\documentclass[12pt]{article}
\usepackage[utf8]{inputenc}
\usepackage[russian]{babel}
\usepackage{epsf}

\begin{document}

\renewcommand{\abstractname}{Abstract}
\renewcommand{\figurename}{Fig.}
\renewcommand{\refname}{References}

\title{
{\bf On the objective origin of the phase transitions and metastability in many-particle systems}}
\author{
{\it A.G. Godizov}\\
{\small {\it Institute for Physics and Power Engineering, 249033 Obninsk, Russia}}\\
{\it A.A. Godizov\thanks{E-mail: anton.godizov@gmail.com}}\\
{\small {\it Institute for High Energy Physics, 142281 Protvino, Russia}}\\
}
\date{}
\maketitle

\begin{abstract}
Equilibrium statistical mechanics is intended to link the microscopic dynamics of particles to the thermodynamic laws for macroscopic quantities. However, the modern 
statistical theory is faced with significant difficulties, as applied to description of the macroscopic properties of real condensed media within wide thermodynamic 
ranges,\linebreak including the vicinities of the phase transition points. A particular problem is the absence of metastable states in the Gibbs statistical mechanics of the 
systems composed of finite number of particles. Nevertheless, accordance between equilibrium statistical mechanics and thermodynamics of condensed media is achievable if to 
take account of the mutual correlation (the feedback) between the microscopic properties of molecules and the\linebreak macrostate of the corresponding medium. This can be 
done via usage of the ``enhanced'' Hamilton operator of the considered many-particle system, which contains some\linebreak temperature-dependent term(s), and the following 
introduction of the generalized\linebreak equilibrium distribution over microstates. For illustration of the reasonableness of the proposed approach (and of its 
availability in practical applications), a cell model of\linebreak melting/crystallization and metastable supercooled liquid for a water-like medium is\linebreak presented.
\end{abstract}

\section*{Introduction}

The modern representations about thermodynamics of many-particle systems were developed in the framework of the subjective approach, wherein the irreversibility (following 
Boltzmann \cite{isihara}) is a consequence of the coarsened perceivence of many-particle systems by the observer, and the relaxation (following Krylov \cite{krylov}) is a 
consequence of the existence of some dynamical unsteadiness in these systems, which results in their mixing up. Keeping in mind these\linebreak conceptions and taking account 
of the short-rangeness of intermolecular forces, one succeeds in construction of the Gibbs distributions \cite{gibbs} which nicely describe the macroscopic properties 
of rarefied gases.

However, any singularities in temperature are absent in the Gibbs statistical sums for the systems composed of finite number of particles (the sums over all the accessible 
microstates). Such statistical sums (partition functions) are one-valued and analytical in temperature\linebreak functions, except at the point $T=0$. Therefore, any attempts 
to provide a statistical description of the macroscopic properties of real condensed media within wide thermodynamic ranges, including the vicinities of the phase transition 
points, face significant difficulties \cite{croxton}. Serious troubles emerge, as well, under the modeling of the condensed phases by the techniques of molecular 
dynamics \cite{vega} (particularly, it regards the combined calculation of both the structural and caloric characteristics \cite{petrenko}). The question {\it ``what 
is the precise nature and origin of the singularities connected with the phase transitions and the critical phenomena?''} \cite{balescu} keeps its actuality. Besides, the 
phase transitions are closely related to the metastable states which always finish their existence via phase transition. Hence, from the practical standpoint, those phase 
transition statistical models seem defective which are exclusive of the metastable states nearby the points of the liquid crystallization or the vapor condensation.

The aim of this paper is, at first, to propose and substantiate the conception that the origin of the phase transitions and metastability in real condensed media is the 
mutual correlation (the feedback) between the particle microscopic properties and the medium macrostate, and, then, to demonstrate, on the example of some simple 
illustrative model, in what way the feedback can lead to the emergence of the metastable states and first-order phase transitions in equilibrium statistical mechanics. In 
conclusion, we propose a general recipe for calculation of the temperature and pressure/density dependences of the macroscopic characteristics of condensed media. 

\section*{The primary conception}

Our reasoning starts from the empirical fact that, in the absence of external influence, the many-particle system metastable states (for instance, supercooled water in a 
test tube) exist, in average, throughout the time intervals much longer than the hydrodynamic relaxation period. Hence, it is sensible to consider them to be equilibrium, 
though unstable with respect to the perturbations stronger than a certain one. (By the way, the notion of metastable state is present in equilibrium thermodynamics, but 
absent in the Gibbs equilibrium formalism.) If to presume the metastable states to be equilibrium, then the main consequence of this presumption, {\it i.e.}, a necessary 
condition for the existence of several equilibrium phase states within the same thermodynamic range, is the temperature many-valuedness of the statistical sum. (Another 
representative example is the existence, in a wide range of temperature and pressure, of two equilibrium phases of carbon: diamond and graphite.)

Keeping in mind that the functional structure of the canonical ensemble is correct (since the rarefied gas properties are described perfectly well by the Gibbs 
distributions), it is reasonable to ask a question: what are the physical prerequisites for the temperature many-valuedness of the statistical sum? The answer is evident: in 
those temperature intervals, wherein the existence of the equilibrium metastable states is possible, the explored system microdynamics should be many-valued in temperature. 
Such a many-valued in temperature structure of microdynamics can be described by means of the ``enhanced'' Hamilton operator which contains some term(s) dependent on 
temperature (the procedure of introduction of such operators is expounded in the following section). According to the consistency principle, the enhanced Hamilton operator 
of an isolated dynamical (deterministic) system, {\it i.e.}, of the finite system whose interaction with the thermostat is negligible objectively, should coincide with 
its usual Hamilton operator. The microdynamical structure of any enhanced Hamilton operator should be different for different equilibrium phases accessible at the 
same temperature. 

Unclosing a self-closed many-particle system by the traditional method, destined just for introduction of the system microstate probabilities, and ignoring any dependence 
of the system quantum state spectrum on the thermostat state, one comes to the situation when even the interaction with the thermostat does not lead to any temperature 
dependence of the system microdynamics.

But if the thermostat is stochastized and the dependence of the system quantum state spectrum on the thermostat influence is significant, then the observer (exploring the 
system) detects that, under the relation $N_S\ll N_W$ (where $N_S$ and $N_W$ are the particle numbers in the system and the thermostat, correspondingly), the system 
perceives the thermostat as a continuous medium. Namely, the system individual particles interact with the stochastic thermostat not in the particle-on-particle manner, but 
in whole. (Indeed, if some medium is stochastized, then it is senseless to consider it discrete in the time scales much longer than the characteristic period between the 
fluctuations.) In other words, in those cases, when the interaction between the explored system and the thermostat is not negligible and the particle-on-particle count of 
this interaction is impossible, the thermostat impact on the system microdynamics should be taken into account in average (through averaging over the thermostat 
microstates). After such an average, the system microdynamics turn out dependent on the thermostat thermodynamic parameters, such as temperature and pressure. If the 
system and the thermostat are in equilibrium, then their temperatures are equal and the system particle concentration is unambiguously determined by the thermostat 
temperature and pressure.

Thus, for such systems, {\it the feedback principle} is valid: 
\begin{itemize}
\item the microdynamics of a many-particle system, which is in equilibrium with its thermostat, correlate with the system macrostate, and this reveals itself in the 
effective dependence of the system quantum state spectrum on the system thermodynamic parameters.
\end{itemize}
The effect of the self-adjustment of the system microdynamics to its macrostate (the feedback) could result in the temperature many-valuedness of the statistical sum and, 
consequently, in the existence of the metastable states and phase transitions. Such systems have possibility to choose among the accessible (at given temperature 
and pressure) types of microdynamics, {\it i.e.}, phases.

For taking the feedback into consideration, it is necessary to keep in mind that atoms and molecules possess electromagnetic structure. The noticeable deformation of 
this structure due to interaction with the environment leads to the emergence of a significant difference between the effective pair interactions in the condensed phases 
and the pair interaction in the rarefied gas state (in other words, in the fundamental level, interaction of molecules in real condensed media is realized via the 
long-range interaction of the molecule charged constituents, and it is not reduced to the molecule pair interaction in vacuum). 

At the end of this section, we would like to emphasize that, for the existence of the feedback, the thermostat objective stochasticity is requisite. The fundamental nature 
of the sources of this stochasticity is not essential for our further reasoning.

In the following sections, after a brief phenomenological analysis of some experimental material, we consider the procedure of introduction of the generalized equilibrium 
distribution over microstates and, then, make the concrete examples which illustrate the above-stated conception about the origin of the phase transitions and 
metastability.

\section*{Substantiation of the feedback principle}

\subsection*{Analysis of the experimental data on the pair distribution functions}

\begin{figure}[ht]
\begin{center}
\epsfxsize=7.5cm\epsfysize=15cm\epsffile{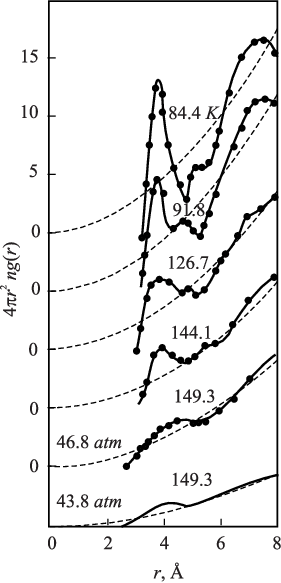}
\end{center}
\vskip -0.5cm
\caption{The particle radial distribution functions in argon at different temperatures \cite{isihara,morgan}.}
\label{argon}
\end{figure}

\begin{figure}[ht]
\begin{center}
\epsfxsize=15cm\epsfysize=6cm\epsffile{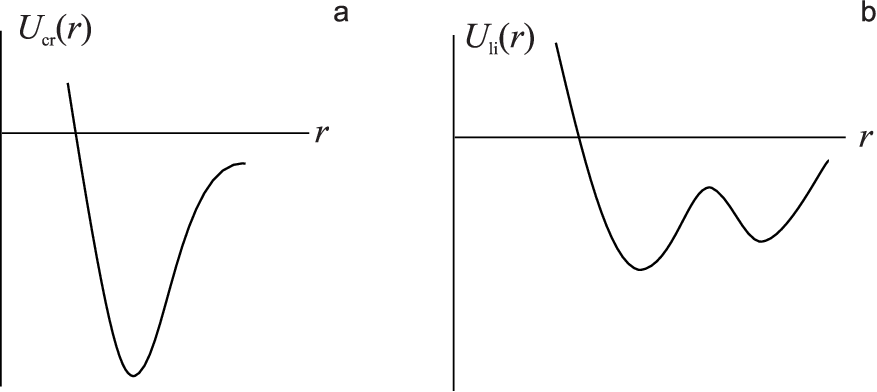}
\end{center}
\vskip -0.5cm
\caption{Qualitative shape of the effective potential energy of the molecule pair interaction: a) in crystal, b) in liquid.}
\label{potent}
\end{figure}

The dependence of the interaction between molecules on the phase state follows from the experimental observations as well. Indeed, the phase transitions are always 
accompanied by the change of the spatial symmetry and structure under melting/crystallization and by the change of the structure under the transition ``liquid-gas''. The 
structure of any molecular system is entirely determined by the interaction between its molecules. Hence, one could assume that, under any first-order phase transition, the 
spatial structure of this interaction ({\it i.e.}, the effective potential energy shape) changes its form sharply. Comparison of the functions of the particle pair 
distribution in crystal and in liquid (for all of substances without any exception) leads to a conclusion \cite{croxton} that in liquid there appears a surplus of molecules 
between the 1st and 2nd coordination spheres (the secondary structure). This is quite distinguishable in the experimental curves of the particle pair distribution in 
liquids \cite{isihara,morgan} (Fig. \ref{argon}). As, in the leading approximation, the pair distribution function is proportional to the factor 
$\exp(-U(|\vec r_1 - \vec r_2|)/k\,T)$ \cite{fisher}, where $U(|\vec r_1 - \vec r_2|)$ is the effective potential energy of the molecule pair interaction, so one could 
infer that, at melting, $U(r)$ undergoes the modification from a function with one minimum to a function with two minima (Fig. \ref{potent}). Physically, this implies that, 
nearby the phase transition point, microdynamics of the same particles in different stable phases ({\it i.e.}, in the crystal and in the liquid) are cardinally different. In 
other words, the structure of the molecule pair interaction, determining the system macrostate, itself depends on this macrostate. 
\vskip -0.5cm

\subsection*{The generalized equilibrium distribution {\it versus} the Gibbs canonical distribution}

Let us proceed to introduction of the generalized equilibrium distribution over microstates. Our consideration will be concentrated on the procedure of quasi-isolation of 
many-particle systems.

We {\it a priori} presume such systems to be immersed into some stochastic (not deterministic) environment $W$ (the thermostat) which possesses a ``very large'' number of 
degrees of freedom (in comparison with the explored system $S$: $N_S\ll N_W$) and stays in thermodynamic equilibrium.

In many-particle systems, the energy spectrum density is closely related to the number of particles in the system: larger number of particles $\rightarrow$ higher density of 
the energy levels. So and thus, at the mutual isolation, the density of the thermostat energy levels is much higher than the density of the system energy levels. 
Correspondingly, even under interaction of $S$ and $W$, the rate of spontaneous quantum transitions (quantum-state reductions) inside the thermostat is much higher than 
inside the system\footnote{In the authors' opinion, as, in the framework of classical theory and quantum mechanics, the matter obeys some time-reversible dynamical 
equations, so the stochasticity sources should be looked for, first of all, in such a property of matter as interconvertibility (the particle creation/annihilation and the 
spontaneous emission/absorption of photons) and in the capacity of composite particles to spontaneous getting excited under collisions, what results in the spontaneous 
changes of the composite particle interactions with their neighbours. Thus, for any many-particle system (even as if entirely isolated), such a thermostat exists as the 
equilibrium background of soft photons.} (if $S$ and $W$ consist of alike particles, then we presume that it is possible, in principle, to divide $S$ and $W$ in space, like 
in the cases of molecular systems):
\begin{equation}
\tau_S\gg\tau_W\,,
\end{equation}
where $\tau_S$ and $\tau_W$ are the characteristic periods between the quantum-state reductions in $S$ and $W$.

Therefore, our main precondition is the possibility of a formal description of the thermostat mixed state by the density matrix 
\begin{equation}
\label{term}
\hat f^W=\sum_mf^W_m(T_W)|w_m><w_m|
\end{equation}
with the coefficients $f^W_m(T_W)$ which represent some equilibrium statistical distribution over the thermostat microstates $|w_m>$. These coefficients depend neither on 
the system $S$ state nor on time, but depend on the thermostat macroscopic characteristics (in particular, on its temperature $T_W$ which can be introduced just formally, 
until the passage to the thermodynamic limit for the thermostat). Namely, we {\it a priori} presume that $\tau_W\to 0$ and consider the thermostat to be an ergodic 
system in the characteristic time scales $dt$ of the observation of the system $S$ microdynamics ($\tau_S\gg dt\gg\tau_W\to 0$).

Between the nearest-in-time quantum-state reductions in $S$, the system state evolution is purely dynamical and determined by the von Neumann equation:
\begin{equation}
\label{fonno}
\frac{\partial\hat f^S}{\partial t} = \frac{1}{i \hbar} \left[\hat H^S + \hat U^{SW}(T_W)\,,\;\hat f^S\right]\,,
\end{equation}
where $\hat f^S = \sum_n f^S_n|s_n><s_n|$ is the system density matrix, and $\hat H^S$ is the Hamilton operator of the isolated system $S$. Operator $\hat U^{SW}(T_W)$, 
obtained via average of the $S$-$W$ interaction operator $\hat U^{SW}$ over the thermostat mixed state ($\hat U^{SW}(T_W)={\rm Sp}\{\hat U^{SW}\hat f^W\}$), 
determines the system interaction with the thermostat average field (this operator depends on the quantum numbers of the system particles and on the thermostat 
thermodynamic parameters). Such an approximation of the system and thermostat interaction is justified, since $\tau_W\to 0$ and, consequently, the system has no time to 
undergo any visible reconfiguration in response to separate fluctuations of the thermostat field (which are caused by the thermostat quantum state reductions). In 
other words, the system individual particles perceive the thermostat as some continuous thermodynamic medium.

The ideology of the mean field approach requires to attribute some part of the $S$-$W$ interaction energy to the thermostat, as the $S$-$W$ interaction mean energy which 
the system is moving about. For this purpose, let us introduce (formally) the mean energy of the $S$-$W$ interaction per particle in the combined system $S\oplus W$: 
\begin{equation}
\overline{U^{SW}} = (N_S+N_W)^{-1}{\rm Sp}\left\{\hat U^{SW}(\hat f^W\otimes\hat f^S)\right\}\,.
\end{equation}
The value of $E^{SW}(T_W)=N_W\overline{U^{SW}}$ should be attributed to the thermostat. Then, operator 
\begin{equation}
\hat H^{S_{eff}}(T_W) = \hat H^S+\lim\limits_{N_W\to\infty}\left[\,\hat U^{SW}(T_W)-E^{SW}(T_W)\right]
\end{equation}
should be defined as the ``enhanced'' Hamilton operator of the effective system $S_{eff}$. It depends on the thermostat thermodynamic parameters. The passage to the 
thermodynamic limit for the thermostat ($N_W\to\infty$, $V_W\to\infty$, $\frac{N_W}{V_W} = const$) is implemented before the introduction of the equilibrium distribution 
over the $S_{eff}$ microstates. 

If the system is in equilibrium with the thermostat (in this paper we restrict ourselves by the equilibrium case), then $T_S=T_W=T$. Besides temperature $T$ and density 
$\rho$ (determined by $T$ and the external pressure), the macrostate of such a system is characterized by the intensive parameters $\{\lambda_a\}$ of external fields. 
Exciting certain degrees of freedom in the system, these fields could cause the system internal reconfiguration and, thus, change the microscopic dynamics in $S_{eff}$ 
(this implies some prospects of the external management of the phases of many-particle systems via the selective influence on those degrees of freedom which are responsible 
for the feedback structure\footnote{In \cite{dubrovich} the issues of some experiment are presented which concerns the impact on supercooled water droplets by the 
radiation of the frequency of the valent oscillations of water molecules. The main (statistically reliable) outcome is that, under the irradiation, the rate of 
crystallization of the droplets increases in several tens of times, in comparison with the crystal formation rate in the shadow zone.}). 

As the Gibbs requirement, relating to the system quasi-isolation, is fulfilled for $S_{eff}$ (its enhanced Hamilton operator $\hat H^{S_{eff}}$ is independent of the 
thermostat dynamical variables), so, following Gibbs \cite{gibbs,balescu}, one can describe the ensemble of such quasi-isolated systems by means of the following 
equilibrium distribution over microstates:
\begin{equation}
\label{effraspr}
f_n^{S_{eff}}(T,\rho,\{\lambda_a\})=Z^{-1}_{S_{eff}}
e^{-\frac{E_n^{S_{eff}}(T,\rho,\{\lambda_a\})}{k\,T}}\,,
\end{equation}
where $E_n^{S_{eff}}(T,\rho,\{\lambda_a\})=<s_n|\hat H^{S_{eff}}(T,\rho,\{\lambda_a\})|s_n>$ is the energy of the effective system $S_{eff}$ in microstate $|s_n>$, and 
$Z_{S_{eff}}$ is the statistical sum for $S_{eff}$. The structure of the potential part of $\hat H^{S_{eff}}(T,\rho,\{\lambda_a\})$ could be not only dependent on $T$, but 
also many-valued in $T$ (as a consequence, the Gibbs measure can).

The requirement $N_S\gg 1$ is used nowhere (under derivation of the Gibbs distribution, it was needed for the system quasi-isolation from the thermostat, under short-range 
intermolecular forces). Thus, the foregoing procedure of effective quasi-isolation (and, consequently, the\linebreak 
generalized equilibrium distribution (\ref{effraspr}) itself) is 
valid, as well, for the systems composed of a few particles. (In the cases of the systems composed of identical bosons or fermions, distribution (\ref{effraspr}) should be 
appropriately modified.)

Note, that the consistency principle is fulfilled: if all the contributions $<s_n|\hat U^{SW}(T_W)|s_n>$ are low enough (or if the dependence of these terms on the 
microstate $|s_n>$ is weak enough), then the $S$-$W$ interaction does not lead to any essential distortion of the system energy spectrum in the vicinity of the 
average value of the system energy. Then, in this vicinity, the spectrum of the system enhanced Hamilton operator $\hat H^{S_{eff}}(T,\rho,\{\lambda_a\})$ reproduces the 
spectrum of its usual Hamilton operator $\hat H^S$ (independent of temperature and density), and distribution (\ref{effraspr}) automatically turns into the Gibbs canonical 
distribution: $f_n^S(T)=Z^{-1}_Se^{-\frac{E_n^S}{k\,T}}$, where $E_n^S=<s_n|\hat H^S|s_n>$. In other words, the class of the many-particle systems described by 
distribution (\ref{effraspr}) is wider than the class of the systems described by the Gibbs distribution (rarefied gases). A particular feature of the latter one is that, 
since the microdynamics of such systems are independent of the thermostat influence, their quantum state spectra turn out independent of temperature, and this results in 
the temperature one-valuedness and analyticity of the corresponding statistical sums.

In the traditional statistical mechanics, for description of condensed media, there are\linebreak exploited some approximations to the spatial structure of particle pair 
interactions, usually obtained through solving the quantum mechanics equations for pairs of atoms or molecules in vacuum. These pair interactions are short-range, and this 
enables to construct the Gibbs ensembles for the quasi-isolated systems whose interaction with the thermostat is neglected. Besides, pair interactions are often designed in 
the form of the sum of attracting and repelling parts. An essential restriction is that the spatial structure of such pair interactions is usually considered to be 
invariable for all of temperatures and phases, {\it i.e.}, it is presumed, by default, that the environment impact on the particle pair interaction is negligible. 
Namely, such particles are implied to be structureless. Just this combination of requirements (the short-rangeness of pair interactions $+$ their independence from the 
environment state), used for explanation of the many-particle system quasi-isolation from the thermostat, is the initial condition for the exclusion of the equilibrium 
metastable macrostates from the Gibbs canonical formalism.

In short, the qualitative difference between the proposed approach (regarding its applicability to description of the macroscopic properties of condensed media) and the 
traditional one is in the following. In the framework of the traditional approach:
\begin{itemize}
\item the requirements of the particle interaction short-rangeness and the largeness of the number of particles (at their fixed concentration) in the 
explored many-particle system are advanced, for the system quasi-isolation from the thermostat;
\item the Gibbs canonical distribution over the system microstates is introduced, while the system energy spectrum is presumed to be independent of the thermostat 
influence;
\item the thermodynamic potentials and other macroscopic characteristics of the system are calculated;
\item the passage to the thermodynamic limit for the system is implemented.
\end{itemize}
In the framework of the proposed approach: 
\begin{itemize}
\item the thermostat impact on the explored system microdynamics is taken into account;
\item the passage to the thermodynamic limit for the thermostat is implemented, {\it i.e.}, the system particle interaction with the stochastic thermostat is 
considered as interaction with continuous medium;
\item the system effective quasi-isolation from the thermostat is implemented by the shift of the system and thermostat interaction energy, via subtraction of its 
mean value;
\item the generalized equilibrium distribution over the system microstates is introduced, while the system energy spectrum is considered dependent on the thermostat 
thermodynamic parameters (in the case of the system and thermostat equilibrium, their thermodynamic parameters are unambiguously interrelated);
\item the thermodynamic potentials and other macroscopic characteristics of the system are calculated.
\end{itemize}

\subsection*{General discussion of the feedback in the context of its impact on the properties of individual molecules in condensed media}

The potential part of operator $\hat H^{S_{eff}}(T,\rho,\{\lambda_a\})$ can be represented as the sum of operators 
$\hat U_{ij}(|\vec r_i - \vec r_j|; T, \rho, \{\lambda_a\})$ of the effective potential energies of pair interaction, which describe the interaction between two particles 
in the presence of the environment whose macroparameters determine the temperature and density variability of $\hat U_{ij}$ (besides the 
distance, $\hat U_{ij}$ can depend on the spatial orientation of the molecules). For instance, the spatial structure of the effective pair interactions in molecular media 
can be calculated through solving the quantum mechanics equations which should contain the medium macroparameters (presetting the environment influence on the charge 
distribution inside the considered molecules). As well, one should keep in mind that the molecule internal reconfiguration, caused by interaction with the environment, 
results not only in some modification of the effective potentials of these molecules, but also in the change of their mass defects, {\it i.e.}, in the change of their 
internal binding energies $\Delta E_j(T,\rho,\{\lambda_a\})$ \cite{hirata} (this change should be explicitly estimated for correct computation of the caloric properties of 
molecular media). In the systems composed of ions and electrons (conductors, plasma), the environment impact on the effective pair interactions results in such a phenomenon 
as Debye screening. The environment ({\it i.e.}, the explored medium) macroscopic properties, in their turn, themselves depend on $\hat U_{ij}$. Thus, from the practical 
point of view, $\hat U_{ij}(|\vec r_i - \vec r_j|; T, \rho, \{\lambda_a\})$ should be interpreted as a self-consistent field. The dependence of $\hat U_{ij}$ on the medium 
macrostate is negligible in certain cases only (within the traditional approach, such an independence is presumed {\it a priori} for any many-particle system). 

The treatment of the molecule pair interaction in the condensed phases as equivalent to the pair interaction of the same molecules in vacuum can lead to unsatisfactory 
outcomes. High concentration of composite particles leads to the noticeable deformation of the molecule internal configurations (in particular, this reveals itself in the 
fact that the average dipole moment of water molecules in the liquid phase is much higher than the dipole moment of water molecule in vacuum \cite{dipol}). 
Thereby, it conditions the environment essential impact on the shape of the effective potential energy of pair interaction. If to compute the spatial structure of the 
effective interaction between two molecules placed into a cavity in some dielectric medium (the cavity sizes are comparable with the sizes of the considered pair), then, for 
high values of the medium electric inductivity, the effective potential energy of the pair interaction could have two minima (by analogy with the potential energy of proton 
under the hydrogen bond inside water dimer in condensate \cite{tapia}). Electric inductivity depends on temperature. Thus, in dielectric media, the molecule effective 
interactions are temperature-dependent as well. In the preface to his monograph \cite{gibbs}, J.W. Gibbs points out that any theory is, certainly, incomplete which ignores 
the deformations of the molecule electromagnetic structure under their interaction: {\it ``In the present state of science, it seems hardly possible to frame a dynamic 
theory of molecular action which shall embrace the phenomena of thermodynamics, of radiation, and of the electrical manifestations which accompany the union of atoms. Yet 
any theory is obviously inadequate which does not take account of all these phenomena.''} Alterable electromagnetic structure of atoms and molecules requires to introduce 
the effective interactions which correlate with the medium macrostate.

Even a simple phenomenological account of the temperature variability of the particle effective interactions significantly advances the practical capabilities on 
description of the thermodynamic properties of condensed media and on the simulation of these media by the techniques of molecular dynamics 
\cite{parsafar,muser,amokrane,hohm,filinov,mishra,robles}. However, for accurate overall thermodynamic description of any real medium, the physical origin of this 
variability should be clarified. The feedback principle makes sensible the modeling of condensed media with the help of the particle effective interactions 
explicitly dependent on the medium macroparameters.

Depending on the feedback structure which prevails in concrete medium (and determines both the phase transition curves and the phase transition types), the parent phase 
outside of the stability region is either absolutely unstable (for instance, the absence of equilibrium superheated crystals) or unstable with respect to the 
perturbations stronger than a certain one (for instance, metastable supercooled liquids). The absence of equilibrium superheated crystals can be explained in the models with 
the feedback causing the modification, with increasing temperature, of the effective potential energy of the molecule pair interaction from a function with one minimum to a 
function with two minima (Fig. \ref{potent}). At the emergence of the second minimum in the effective potential energy shape, an immediate loss of the crystal phase 
symmetry occurs (the absolute instability). Inversely, under cooling the liquid down, the statistical disarrangement, having led to the formation of the second minimum, 
prevents its disappearance, since a synchronous passage over potential barrier of the molecules in some macroscopic region is needed (the stable phase kernel).

We would like to emphasize once again: for the presence of the equilibrium metastable states in the statistical mechanics of any many-particle system, the potential part 
of the corresponding (enhanced) Hamilton operator should {\it indispensably depend on temperature and be many-valued in temperature in the ranges of the metastable state 
existence}. Otherwise, the system microdynamics do not correlate with its macrostate, and, then, the existence of any equilibrium metastable phase is impossible.

Below we consider a cell model for a water-like medium, wherein the feedback is embodied via the molecule effective interaction dependent on the medium dielectric 
properties which are determined, in their turn, by the effective potential energy shape. This model reveals a first-order phase transition and an equilibrium 
metastable state of supercooled liquid.

\section*{Illustrative examples of the feedback mechanism}

\subsection*{A mechanical model of metastable state}

\begin{figure}[ht]
\epsfxsize=16.5cm\epsfysize=6cm\epsffile{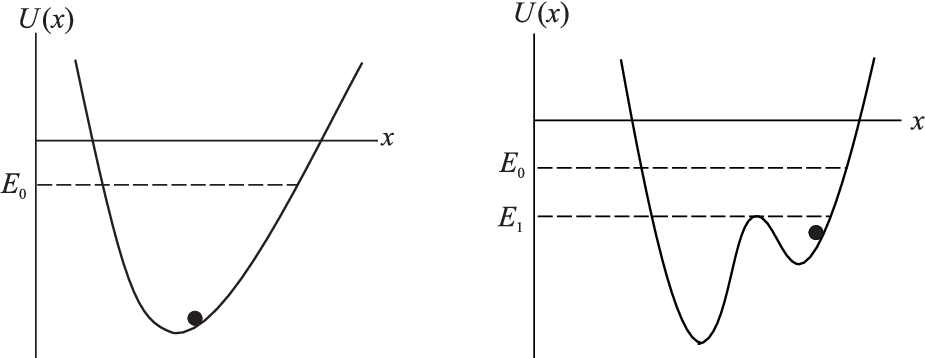}
\caption{A mechanical model of metastable state.}
\label{mechan}
\end{figure}

Let us consider a potential well with elastic walls, in which there rolls frictionless a heavy ball (Fig. \ref{mechan}, on the left). By the instrumentality of some 
mechanical appliances, the well is given a shape with two dips (Fig. \ref{mechan}, on the right). The system state is determined by the ball total energy $E_0$ and the 
range of accessible positions. At $E_0>E_1$, all the positions with $U(x)<E_0$ are accessible to the ball. At $E_0<E_1$, the ball can be either within the primary dip or 
within that one which is formed by the appliances, and, in the absence of fluctuations, the ball is not able to transit from one dip into another one.

In the presence of fluctuations, the ball randomly changes its microstate from time to time. One should not consider the ball macrostate within the deeper dip to be more 
stable in comparison with the macrostate within another one, since the relative time spent by the ball within any dip depends on the fluctuation spectrum. A quite different 
situation is if, at the ball staying in the secondary dip, to take away the appliances which provide the existence of this dip. In the case of the lack of the elastic 
energy of the well wall, the ball will keep staying in this dip until leaving it under the action of fluctuations. Just this macrostate should be considered metastable, 
since, upon the ball transition into the primary dip, the well restores its initial form, and the ball is not able to return into the disappeared dip. 

The given example allows to formulate, in rather general form, the conditions for the existence of the metastable states. The metastable states emerge in the systems 
with fluctuations and self-action via the environment field of response. The presence of fluctuations is necessary but not sufficient for the metastability. 
The sufficient condition is the system self-action via the environment field of response, keeping up the current macrostate just until the system transition into another 
one. Upon the system has passed into the stable state, the metastable phase collapses and appears no more. 

\subsection*{A fixed-volume cell model of melting/crystallization and metastable supercooled liquid for a water-like medium}

Our reasoning could be considered purely speculative, if the explicit demonstration of how the feedback works in condensed media were missing in, though the 
above-presented extensive part of the paper is absolutely essential for understanding that the exhibited below model is not reduced to some mathematical trick, but serves 
as the simplest illustration of the fundamental physical mechanism which leads to the emergence of both the equilibrium metastable states and the phase transitions in 
many-particle systems.

Let us consider the feedback phenomenon on the example of a fixed-volume cell model. Our aim is not to provide an accurate overall thermodynamic description of any 
real dielectric medium (this is a very laborious goal and it should be the aim of a separate work), but just to illustrate the prospects of taking account of the feedback 
for the modeling of the interactions of particles. Therefore, we simplify our consideration, to make more transparent the origin of the major effects (the 
metastability and the first-order phase transition). The elementary cell itself is fixed to be spherically symmetric with radius $R_c=2.76\,\dot A$ (for real media, the 
hydrogen bond directionality should be taken into account). As well, for simplicity, we ignore the dependence of the molecule internal binding energy (mass defect) on the 
environment macrostate and consider the cell molecule as a classical point-like particle with the mass equal to $3\cdot 10^{-23}g$.

In real materials (in particular, in water), under the hydrogen bond (HB) formation, the charge densities inside the molecules are redistributed, and the corresponding pair 
of molecules acquires an extra dipole moment. In crystal of ice, any molecule with the complete set of HBs is electrically neutral. Under breaking the HB, each of the 
molecules acquires an effective charge $\pm Q_{eff}(r)$ ($r$ is the distance from the cell center). The effective interaction of such molecules includes not only the 
interaction with their neighbours, but also the interaction of charges $\pm Q_{eff}(r)$ with the continuous dielectric environment. Here we make an extra simplification by 
presuming that there are only two types of molecules in the considered medium: the molecules with the complete set of HBs and the molecules with one HB broken. 

The effective potential energy of the cell particle is introduced as 
\begin{equation}
U_{eff}(r,T)=U_0(r)+U_\epsilon(r,T)\,.
\label{potwat}
\end{equation}
Function $U_0(r)$ is preset with the help of some parametric (test) approximation to the attractive part of the interaction between two water molecules in vacuum, 
estimated numerically by the techniques of quantum mechanics \cite{stillinger}:
\begin{equation}
\label{poten0}
U_0(r)=\left\{
\begin{array}{lcl}
\hspace*{2.45cm}\alpha(r-a)\,,\;\;\;\;\;0\le r\le R_l\\
\beta\{\sqrt{r-b}-\sqrt{R_c-b}\}\,,\;\;\;\;R_l< r\le R_c\\
\hspace*{3.5cm}\infty\,,\;\;\;\;R_c<r<\infty
\end{array}\,,
\right.
\end{equation}
where $\alpha=3.36\cdot 10^{-5}\,erg\cdot cm^{-1}$, $a=1.25\,\dot A$, $R_l=0.75\,\dot A$, $\beta=1.364\cdot 10^{-9}\,erg\cdot cm^{-1/2}$, and $b=0.71\,\dot A$. 
Contribution $U_\epsilon(r,T)$ is chosen to have the following form:
\begin{equation}
\label{potend}
U_\epsilon(r,T)=\left\{
\begin{array}{lcl}
-[\gamma+\lambda N_1(T)]r\;\;\;+\Delta(T)\,,\;\;\;\;\;0\le r\le R_H\\
-[\gamma+\lambda N_1(T)]R_H+\Delta(T)\,,\;\;\;\;R_H\le r<\infty
\end{array}\,,
\right.
\end{equation}
where $0\le r\le R_H=2\,\dot A$ is the HB existence region, $N_1(T)$ is the specific number of the molecules which have broken one HB, parameter 
$\gamma=3.19\cdot 10^{-6}\,erg\cdot cm^{-1}$ is related to the medium dielectric properties at $T=0$, parameter $\lambda = 3.46\cdot 10^{-28}\,erg\cdot g\cdot cm^{-1}$ 
characterizes the variability of the molecule polarization properties under breaking the HB, and $\Delta(T)$ is an additive constant related to the mean interaction with 
the environment (the magnitude of the deviation of the mean interaction energy from its value at $T=0$).

The effective potential energy $U_{eff}$ of the cell molecule depends on the fraction of the molecules which have broken the HB (since the medium electric 
inductivity depends on this number), and number $N_1(T)$ itself depends on $U_{eff}$. If the fraction of the molecules having broken the HB is higher than some critical 
value, $N_1(T)>N_{cr}=\frac{\beta}{2\lambda\sqrt{R_H-b}}-\frac{\gamma}{\lambda}$, then there emerges an extra minimum in the curve of $U_{eff}$. Assuming that the particles 
distributed over elementary cells obey Boltzmann statistics, we come to the equation for $N_1(T)$:
\begin{equation}
\label{chismol}
N_1(T)=N\int_{R_H;\,R^*(T)}^{R_c}e^{-\frac{U_{eff}(r,T)}{k\,T}}r^2dr\left[\int_0^{R_c}e^{-\frac{U_{eff}(r,T)}{k\,T}}r^2dr\right]^{-1}\,,
\end{equation}
where $N\approx 3.33\cdot 10^{22}g^{-1}$ is the total specific number of molecules, and $R^*(T)=b+\frac{\beta^2}{4[\gamma+\lambda N_1(T)]^2}$ is the distance (from the cell 
center) at which $\frac{\partial U_{eff}(r,T)}{\partial r}$ vanishes (at $N_1(T)>N_{cr}$), {\it i.e.}, it indicates the boundary between the elementary cell regions with the 
molecules having the HB broken (the region $R^*(T)\le r\le R_c$) or existing (the region $0\le r\le R^*(T)$).

Thus, the quintessence of the model is the mutual correlation (the feedback) between the effective microscopic potential and some macroscopic quantity obtained through 
average over the statistical ensemble.

\begin{figure}[ht]
\vskip -0.2cm
\begin{center}
\epsfxsize=7.7cm\epsfysize=7.7cm\epsffile{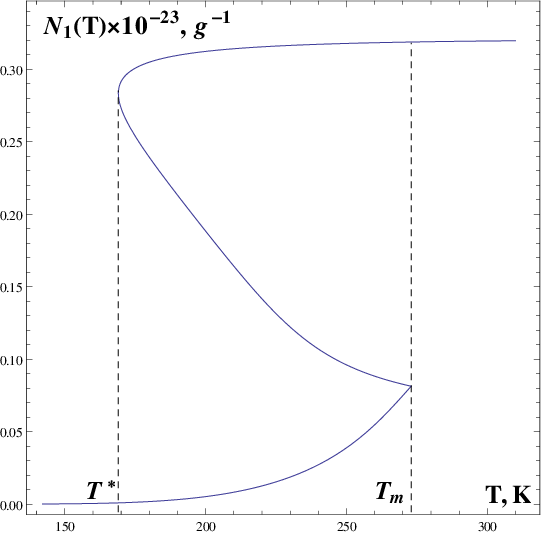}
\end{center}
\vskip -0.5cm
\caption{The temperature dependence of the specific number of the molecules which have broken the HB.}
\label{chis}
\end{figure}

The numerical solution of (\ref{chismol}) is presented in Fig. \ref{chis}.
\vskip 0.2cm

\begin{figure}[ht]
\vskip -0.2cm
\begin{center}
\epsfxsize=7.7cm\epsfysize=7.7cm\epsffile{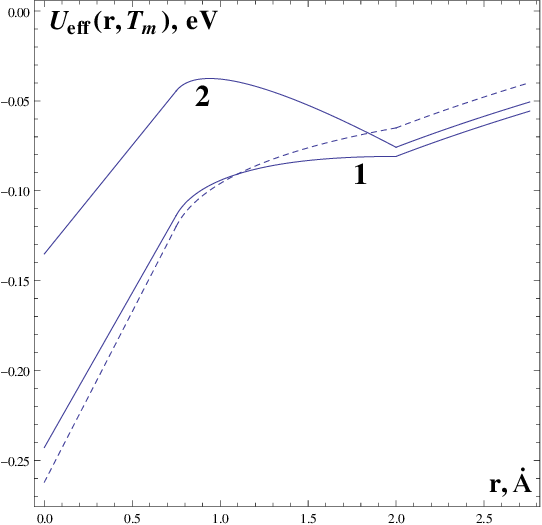}
\end{center}
\vskip -0.5cm
\caption{Two types of the molecule effective potential energy, realized in the elementary cell at the melting temperature $T_m\approx 273\,K$ (curve 1 corresponds to the 
crystal, curve 2 --- to the liquid). The dashed line is the effective potential energy in the crystal at $T=0$.}
\label{ueff}
\end{figure}

In Fig. \ref{ueff} there are depicted two types of the molecule effective potential energy, realized in the elementary cell at the melting temperature $T_m\approx 273\,K$ 
(curve 1 corresponds to the crystal, curve 2 --- to the liquid). The latent heat of melting $\Lambda(T)$ is proportional to the difference between the mean values of 
$U_{eff}$ in the liquid and in the crystal at temperature $T$ (let us remind that in our illustrative model we ignore the dependence of the molecule internal binding energy 
on the environment macrostate). In the considered model: $\Lambda(T_m)\approx 3.26\cdot 10^9\,erg/g$.

\begin{figure}[ht]
\hskip 0.1cm
\epsfxsize=7.7cm\epsfysize=7.7cm\epsffile{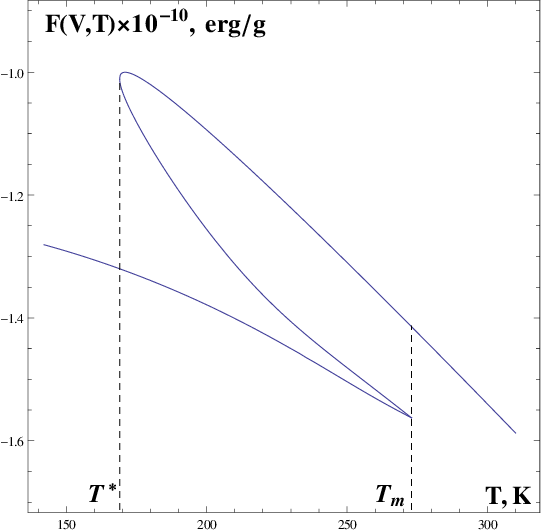}
\vskip -7.75cm
\hskip 9cm
\epsfxsize=7.7cm\epsfysize=7.7cm\epsffile{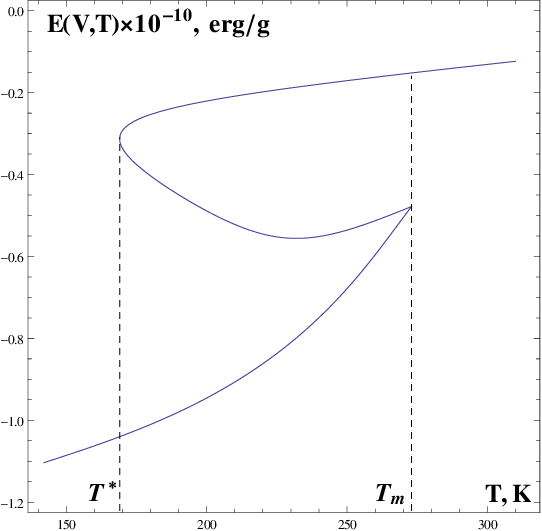}
\vskip -0.2cm
\caption{The temperature dependences of the specific free energy $F(V,T)$ and the specific internal energy $E(V,T)$ at the fixed volume.}
\label{ener}
\end{figure}

In Fig. \ref{ener} there are presented the temperature dependences of the specific free energy $F(V,T)$ and the specific internal energy $E(V,T)$ at the fixed volume. Within 
the interval $T^*<T<T_m$, the existence of the supercooled liquid state is possible ($T^*\approx 169\,K$ is the temperature of the liquid maximal supercooling). 
Varying the cell volume, it is feasible to compute the surfaces of the thermodynamic potentials in the space of variables $V$ and $T$, with the characteristic folds in the 
vicinity of the melting point, as in catastrophe theory \cite{poston}.

\begin{figure}[ht]
\begin{center}
\epsfxsize=7.7cm\epsfysize=7.7cm\epsffile{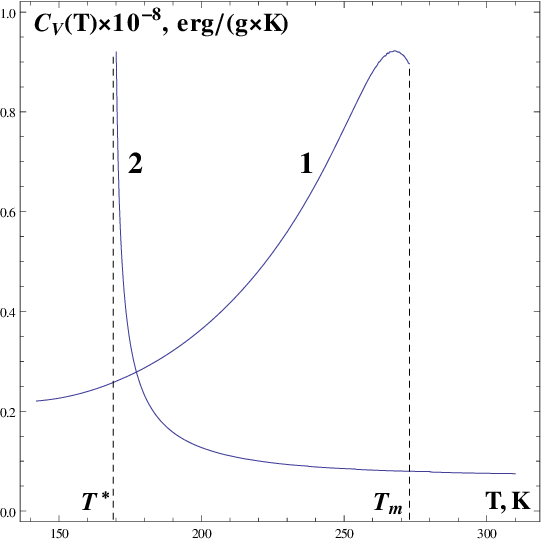}
\end{center}
\vskip -0.5cm
\caption{The temperature dependence of the fixed-volume specific heat capacity: curve 1 corresponds to heating the crystal, curve 2 --- to cooling the liquid.}
\label{capa}
\end{figure}

In Fig. \ref{capa} there is presented the temperature dependence of the fixed-volume specific heat capacity (curve 1 corresponds to heating the crystal, curve 2 --- to 
cooling the liquid). The specific heat capacity $C_V(T)$ exhibits a hysteresis behavior, since, under changing temperature, the system evolves in accordance with the 
Le Chatelier principle.

\newpage

\vskip -0.5cm
\vskip 0.3cm

The superheated crystal state is absent in the model.

Resuming the premises, we conclude that the model reveals a first-order phase transition, an equilibrium metastable state of the liquid phase, and the asymmetry of the 
transition from the ordered phase to the disordered one and back. The phase transition emerges at the specific volume fixed. All these effects take place due to 
contribution (\ref{potend}) into the effective potential energy (\ref{potwat}) of the cell particle. The absence of such a contribution (responsible for the feedback) 
automatically results in the absence of all the above-mentioned phenomena.

In spite of its simplicity and roughness, the presented statistical model provides a qualitative equilibrium description of both metastability and melting/crystallization 
and can serve as the simplest illustration of the necessity to take the feedback into consideration for the modeling of condensed media.

\section*{Conclusions}

Interaction between the explored system and the thermostat (which, due to its principal stochasticity, should be considered as a continuous medium) results in the 
dependence of the system quantum state spectrum on the thermostat macroparameters. In some cases this dependence is negligible, but mostly it is not. If the system is in 
equilibrium with the thermostat, then the values of its thermodynamic parameters are unambiguously interrelated with the thermostat thermodynamic characteristics, and, 
thereby, the system quantum state spectrum turns out effectively dependent on the system macrostate. Such a correlation of the system microdynamics with its macrostate (the 
feedback) can be described via introduction of the system enhanced Hamilton operator whose potential part is dependent on the system\linebreak 
macroparameters. The feedback can lead 
to various regimes of the system microdynamics at the same temperature, {\it i.e.}, to possibility of the system existence in different phases. Just the feedback is the 
underlying cause of the phase transitions and metastability in many-particle systems. Taking account of the feedback significantly advances the capabilities of equilibrium 
statistical mechanics on computation of the thermodynamic properties of condensed media (including the phase transition curves in the spaces of thermodynamic parameters).

The main conclusion is that 
\begin{itemize}
\item for construction of adequate models of condensed media, it is {\it inevitable} to take account of the dependence of the internal structure of atoms and molecules 
(and, consequently, of the spatial structure of the particle effective interactions) on the macroscopic properties of the explored medium.
\end{itemize}

The medium macroparameters, in their turn, are entirely determined by the properties of the particles which compose the medium. Hence, the reliable way to solve the problem 
of the phase transitions and metastable states in real condensed media and to provide an accurate {\it ab initio} computation of the temperature and pressure/density 
dependences of the macroscopic characteristics of various materials (such as electric inductivity, specific heat capacity, viscosity, {\it etc.}) is in the {\it combined} 
(it is important!) determination of the following intercorrelating\linebreak quantities:
\begin{itemize}
\item the spatial structure of the particle effective interactions, {\it i.e.}, the potential energy shape, which is explicitly dependent on the explored medium 
macroparameters, 
\item the particle distribution functions (dependent on the particle effective interactions), 
\item the average value of the particle internal binding energy (for composite particles), 
\item the medium macroparameters themselves (determined by the particle distribution\linebreak functions and the spatial structure of the particle microscopic interactions).
\end{itemize}
Concrete realizations of this rather general recipe, regarding its employment in materials science or chemical physics, are, certainly, different for different media.

Point out, that, following carefully this recipe in applications to real condensed media, one needn't to introduce any free phenomenological parameters. All the 
macroparameters, which preset the explored medium impact on the individual particle properties (and, thus, determine the temperature and pressure/density dependences of the 
effective pair interactions), should be calculated in the framework of the above-stated self-consistent problem. Consequently, it is inevitable for all of the models of real 
condensed media, which are based on the proposed approach, to be explicitly falsifiable (in the sense of K.R. Popper's terminology). Under\linebreak simplified consideration, 
one could use some approximations to microscopic interactions with free parameters to be fitted to experimental data. However, these test parametrizations should provide 
a correct representation of the self-consistent thermodynamic evolution of both the individual particle properties and the medium macroscopic characteristics.

In the very end, we would like to note that, without appealing to the feedback principle, it is difficult to explain the origin of the directivity of the evolution of 
complex systems (including various ecosystems \cite{capitan}). The feedback principle enables to understand, in what way a randomly appeared macroscopic feature is being 
carried into the microscopic level, consolidating itself, and developing. If it were no feedback, the emergence and decay of such features would be equiprobable, and 
it would be no singled out direction of the evolution. The feedback violates this symmetry (``the arrow of time'', autocatalysis). The microscopic-level accumulation of 
the feature ($+$ the gradual thermodynamic changing of microdynamics) could, at last, bring the system into some region of instability and, consequently, to a phase 
transition. Quantity converts itself into quality. This is possible only in the systems with the feedback.

\vskip 0.3cm

{\bf Acknowledgements.} The authors are indebted to P.N. Svirkunov, B.V. Zatolokine, V.A. Petrov, S.M. Klishevich, G.P. Pronko, and V.V. Kiselev for discussions and useful 
criticism.

\end{document}